# Is Large Language Model All You Need to Predict the Synthesizability and Precursors of Crystal Structures?

Zhilong Song[1], Shuaihua Lu[1], Minggang Ju[1], Qionghua Zhou[1,2,*], and Jinlan Wang[1,2,*]

[1]Key Laboratory of Quantum Materials and Devices of Ministry of Education, School of Physics, Southeast University, Nanjing 21189, China

[2] Suzhou Laboratory, Suzhou, China

Accessing the synthesizability of crystal structures is pivotal for advancing the practical application of theoretical material structures designed by machine learning or high-throughput screening. However, a significant gap exists between the actual synthesizability and thermodynamic or kinetic stability, which is commonly used for screening theoretical structures for experiments. To address this, we develop the Crystal Synthesis Large Language Models (CSLLM) framework, which includes three LLMs for predicting the synthesizability, synthesis methods, and precursors. We create a comprehensive synthesizability dataset including 140,120 crystal structures and develop an efficient text representation method for crystal structures to fine-tune the LLMs. The Synthesizability LLM achieves a remarkable 98.6% accuracy, significantly outperforming traditional synthesizability screening based on thermodynamic and kinetic stability by 106.1% and 44.5%, respectively. The Methods LLM achieves a classification accuracy of 91.02%, and the Precursors LLM has an 80.2% success rate in predicting synthesis precursors. Furthermore, we develop a user-friendly graphical interface that enables automatic predictions of synthesizability and precursors from uploaded crystal structure files. Through these contributions, CSLLM bridges the gap between theoretical material design and experimental synthesis, paving the way for the rapid discovery of novel and synthesizable functional materials.

## 1. Introduction

To address the high cost of traditional trial-and-error materials design, numerous theoretical approaches have been employed such as high-throughput (HT) screening based on density functional theory (DFT)[1–3] and data-driven machine learning (ML)[4–9]. HT allow for atomic-level simulations to identify materials with desired properties, while ML provides rapid and accurate property predictions, greatly accelerating the discovery process. However, HT and ML approaches are limited to known structures in existing databases, while inverse design using generative models offers a promising solution for discovering novel material structures[10–14], such as semiconductors, zeolites and metal-organic frameworks[15–21]. Despite these advancements, a common challenge lies in these theoretical methods—the synthesizability of the predicted material structures. Regardless of the efficiency of the method used, the predictions are meaningless if the predicted material structures cannot be synthesized.

The common methods used to screen synthesizable material structures include evaluating thermodynamic formation energies or energy above the convex hull[14,22,23]. However, despite being predicted to have favorable formation energies, many material structures fail to be synthesized. Conversely, many useful metastable structures were successfully synthesized even though their formation energies were not at the convex hull minimum[24]. Another method is assessing kinetic stability, indicated by the absence of imaginary frequencies in phonon spectra[25]. Yet, many material structures with imaginary frequencies in DFT-calculated phonon spectra can still be synthesized experimentally[26]. Furthermore, material synthesis, however, is a complex process influenced by synthesis precursors and conditions[27], in addition to thermodynamic and kinetic stability. Thus, there is an urgent need for a method that can accurately predict the synthesizability of crystal structures and their precursors.

Recent advances in large language models (LLMs), such as OpenAI's ChatGPT[28]. and open-source LLMs like Meta's LLaMA[29,30], have revolutionized various research domains. In materials science, LLMs assist in predicting material properties[31], optimizing experimental workflows[32], and synthesizing scientific literature[33]. For example, fine-tuned GPT-3 can perform comparably to or even outperform conventional machine learning models in various chemistry and materials science tasks, especially with small datasets[31]. The Coscientist, powered by GPT-4, can autonomously design and execute complex experiments, significantly accelerating diverse scientific research tasks [32]. These LLMs demonstrate exceptional capabilities to learn from textual data due to their extensive architectures and vast training datasets. Thus, LLMs hold great potential to predict the synthesizability and precursors of theoretical crystal structures using material structure data in large-scale databases and precursor recipes in literature.

In this work, we developed an LLMs framework called Crystal Synthesis Large Language Models (CSLLM). The framework includes a Synthesizability LLM for predicting the synthesizability of crystal structures, a Methods LLM for classifying synthesis methods (solid-state or solution), and a Precursors LLM for predicting synthesis precursors. 70,120 synthesizable crystal structures were selected from the Inorganic Crystal Structure Database (ICSD)[34] and a pre-trained PU learning model[35] was used to screen 80,000 non-synthesizable structures from 1,401,562 theoretical structures, creating a balanced dataset for synthesizability prediction. We also developed an efficient material structure text representation method—material string, which includes information on space group, lattice, composition, and Wyckoff positions. After fine-tuning with 140,120 material string-synthesizability label mapping data, the Synthesizability LLM achieved 98.6% accuracy on the test set, surpassing DFT methods based on thermodynamic and kinetic stability by 106.1% and 44.5%, respectively. The Precursors and Methods LLMs, fine-tuned with chemical formula-precursor mapping data from literature, achieved classification accuracies of 91.02% and precursor prediction success rates of 80.2%, respectively. To enhance reliability and applicability, we calculated reaction energies to ensure thermodynamic feasibility and stability, and conducted a combinatorial analysis of initial precursor suggestions to generate more potential precursors. Finally, we developed a graphical interface for CSLLM (Figure S1), allowing users to upload crystal structure files or input material strings to automatically generate synthesizability and precursor predictions. This groundbreaking framework addresses the critical challenge of synthesizability, bridging the gap between theoretical predictions and practical synthesis, and thereby paving the way for the efficient discovery and development of novel materials.

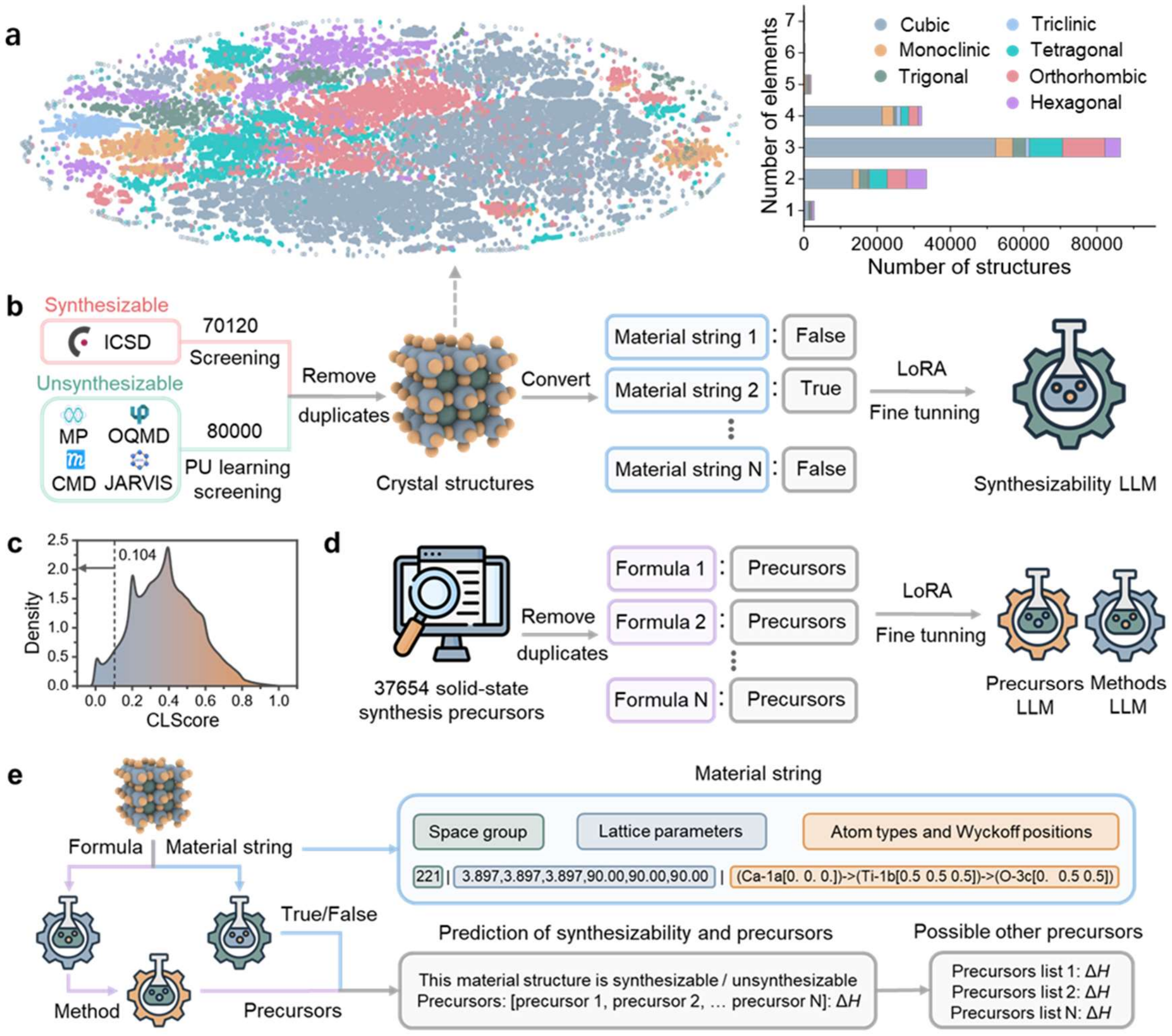

**Figure 1. (a)** T-SNE visualization of 140,120 structures along with statistical data on crystal systems and the number of elements. **(b)** Training process of the Synthesizability LLM. A balanced dataset is constructed using a PU learning model, converted to material strings in batches, and then fine-tuned with LoRA. **(c)** CLScore distribution for 1,401,562 structures and the CLScore range for filtering out non-synthesizable structures. **(d)** Training process for the Precursors and Methods LLMs. Data pairs of 37,654 chemical formulas with their precursors and synthesis methods are collected from the literature and fine-tuned with LoRA. **(e)** Overall workflow for predicting the synthesizability and recommending precursors using the LLM. Crystal structures are first converted into material strings. Their synthesizability are then predicted by the Synthesizability LLM. Based on the chemical formulas, the Precursors and Methods LLMs provide a batch of potential precursors along with their reaction energies.

## 2. Results and discussion

### 2.1. Balanced and comprehensive dataset for predicting synthesizability

To develop a robust LLM for predicting the synthesizability of inorganic crystal structures, it is imperative to establish a comprehensive dataset comprising synthesizable (positive examples) and non-synthesizable (negative examples) materials. The ICSD[34] serves as a reliable source of experimentally validated material structures, all of which are confirmed to be synthesizable. For our study, we meticulously selected 70,120 crystal structures from the Inorganic Crystal Structure Database (ICSD), each containing fewer than 40 atoms and fewer than 7 components, to serve as our positive examples. Notably, disordered structures within the ICSD were excluded from our selection criteria, as our objective focuses on the prediction of the synthesizability of ordered crystal structures. To construct a set of negative examples, we employed a pre-trained Positive-Unlabeled (PU) learning model developed by Jang *et al*[35]. This model generates a CLscore for each structure, with scores below 0.5 indicative of non-synthesizability. This enabled us to discern non-synthesizable structures from a vast pool of 1,401,562 crystal structures sourced from the Materials Project (MP)[36], Crystallography Open Database (COD)[37], Open Quantum Materials Database (OQMD)[38], and Joint Automated Repository for Various Integrated Simulations (JARVIS)[39]. In order to build a balanced dataset, we calculated the CLscores for all 1,401,562 structures and selected the 80,000 structures with the lowest CLscores (i.e., CLscore < 0.104) as our non-synthesizable examples (Figure 1c). Furthermore, we computed the CLscores for our 70,120 positive examples, revealing that 98.3% of these structures had CLscores greater than 0.104, thereby affirming the validity of our CLscore threshold.

All 140,120 crystal structures are visualized using t-SNE[40], covering the seven crystal systems: cubic, hexagonal, tetragonal, orthorhombic, monoclinic, triclinic, and trigonal. These were represented by gray, purple, cyan, red, orange, blue, and green dots, respectively, with the Cubic system being the most prevalent (Figure 1a). Additionally, our dataset includes crystal structures with 1 to 7 elements, predominantly featuring 2 to 4 elements, and covers elements 1 to 94 from the periodic table, excluding elements 85 and 87 (Figure S2). This demonstrates that our dataset is both balanced and comprehensive, laying a solid foundation for the subsequent training of a high-fidelity LLM for predicting the synthesizability of inorganic crystal structures.

## 2.2. Textualization of material structures and fine-tuning of synthesizability LLM

Given that LLMs process text inputs, it is essential to represent material structures in the simplest reversible text format that includes comprehensive information on the lattice, composition, atomic coordinates, and symmetry. The most common text representations of crystal structures are the CIF format[41] and the POSCAR format used by the Vienna Ab initio Simulation Package (VASP)[42]. These formats provide detailed information on the lattice, composition, and atomic coordinates but include redundant information. For example, multiple atomic coordinates at the same Wyckoff position can be inferred from one atomic coordinate along with the space group and Wyckoff position symbols[43]. Therefore, it is unnecessary to list all the atomic coordinates within the cell. Additionally, POSCAR format lacks symmetry-related information, although it is more concise than CIF. To address these issues, we propose a text representation for crystal structures named material string: SP | a, b, c, α, β, γ | ($AS_1$-$WS_1$[$WP_1$])->($AS_2$-$WS_2$[$WP_2$])->($AS_N$-$WS_N$[$WP_N$]), where the SP, (a, b, c, α, β, γ), AS, WS and WP are the space group number, lattice constants, atom symbol, Wyckoff position symbol and Wyckoff position of a crystal structure, respectively. For instance, the material string of the $CaTiO_3$ structure (mp-4019) from MP database is "221 | 3.897.3.897.3.897.90.00.90.00.90.00 | (Ca-1a[0.0.0.])->(Ti-1b[0.5 0.5 0.5])->(O-3c[0.5 0.5 0.5])"(Figure 1e). Notably, crystal structures are first converted into primitive cells before converting to material strings.

The material string not only contains all the crystallographic information but is also as concise as possible. For example, the converted $CaTiO_3$ structure has a string length of only 102 characters, significantly shorter compared to 1817 characters in CIF format and 1644 characters in POSCAR format. This high-information-density text is beneficial for improving the performance of fine-tuning LLMs[44]. Thus, we converted 140,120 unique crystal structures and their synthesizability labels into material strings—True/False data pairs, then split them into training and testing sets in a 9:1 ratio. Considering the high cost of fully fine-tuning LLMs, we employed the LoRA technique (Low-Rank Adaptation)[45], which allows efficient fine-tuning by optimizing only a small subset of parameters, to fine-tune the LLaMA-7B model.

The fine-tuned synthesizability LLM achieved remarkably high accuracy on the test set, with only 88 incorrect predictions out of 7,529 synthesizable crystal structures (Figure 2a). To demonstrate that the outstanding ability of LLM to distinguish synthesizable crystal structures is a result of our fine-tuning rather than inherent to the model, we trained the synthesizability LLM with 0, 1,000, 10,000, 50,000, and 135,108 data pairs. As shown in Figure 2b, the recall (the proportion of actual positives correctly identified) and fallout (the proportion of actual negatives incorrectly identified as positive) of the untuned LLM were only 53.1% and 49.0%, respectively, which is almost equivalent to random guessing. After fine-tuning with 1,000 data points, the recall and fallout of synthesizability LLM improved to 79.8% and 9.36%, demonstrating the powerful ability of LLMs to learn patterns from textual data. Ultimately, the accuracy, precision, recall, and F1 score reached 98.6%, 98.8%, 98.4%, and 98.6%, respectively. The detailed testing performance of the five LLMs in Figure 2b are shown in Table S1 and Figure S3.

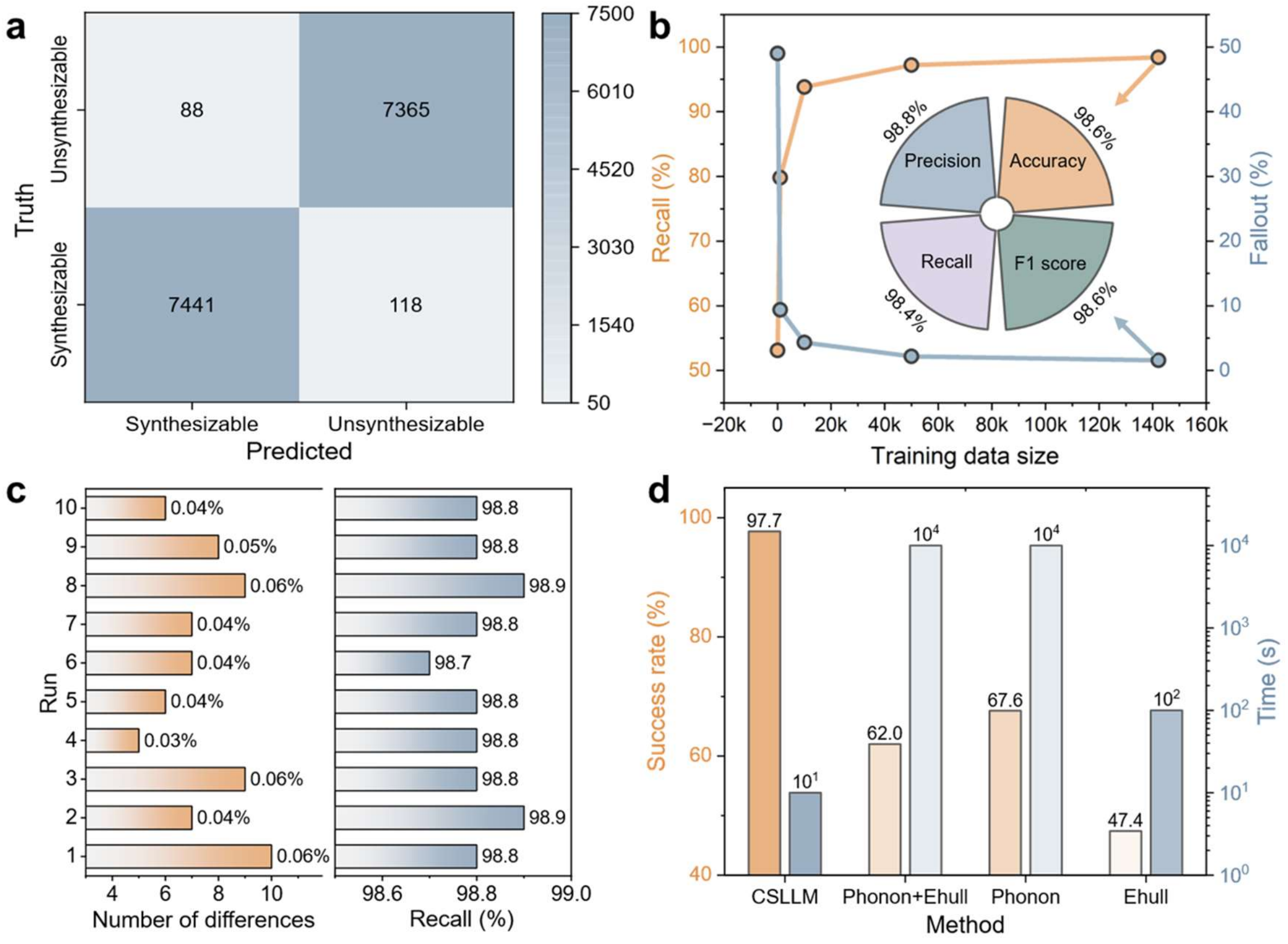

**Figure 2. (a)** Confusion matrix of the Synthesizability LLM on the test set. **(b)** Recall and fallout of the Synthesizability LLM under different amounts of training data. **(c)** Variations in results and recall of Synthesizability LLM for 10 additional test set inferences compared to the initial results. **(d)** Comparison of the accuracy and runtime of synthesizability predictions for crystal structures between the Synthesizability LLM and DFT computational methods.

However, due to variations in random seeds, identical prompts input into the LLM often yield different results. To demonstrate that the synthesizability LLM can reliably and accurately predict the synthesizability of crystal structures, we conducted an additional 10 inference runs on the test set. As shown in Figure 2c, the largest discrepancy in synthesizability predictions compared to the first run was only 10 instances, representing 0.06% of the total. The recall rate varied by no more than 0.2%, affirming the stability of the synthesizability LLM.

Building on this stability, we further compared the synthesizability LLM with three DFT-based methods commonly used for screening experimentally synthesizable material structures: (i) energy above hull (Ehull) = 0 eV/atom, (ii) the lowest frequency of the phonon spectrum (Phonon) = 0 THz, and (iii) both $E_{hull}$ = 0 eV/atom and Phonon = 0 THz. This comparison utilized data from 1512 structures in the MP database for which phonon spectrum calculations had been conducted. The $E_{hull}$, the lowest phonon frequency, and labels indicating experimental synthesis of these structures are collected. We then performed inferences on these structures using the synthesizability LLM and measured accuracy based on the proportion of experimentally synthesized structures that met the criteria of the aforementioned methods. As shown in Figure 2d, the synthesizability LLM achieved a significantly higher accuracy of 97.7% compared to the other methods, while also reducing runtime by 1-3 orders of magnitude relative to DFT methods, demonstrating its capability to accurately and rapidly predict the synthesizability of crystal materials.

## 2.3. Fine-tuning of Methods and Precursors LLM

After identifying the synthesizable material structures, precursors need to be selected before experimental synthesis. Common synthesis methods include the solid-state method and the solution method, each using different precursors. To address this, we integrated data from 19,488 solid-state synthesis precursors collected by Kononova *et al.*[46] and 35,675 solution synthesis precursors collected by *Wang et al.*[47] Additionally, we extracted 6136 data using these two syntheses methods from 2021 to 2024, utilizing scripts provided by Kononova *et al* [46]. We first constructed structure-precursor data pairs from this dataset and converted all structures into material strings. Since not all entries contained material structure information, we ultimately retained only 3650 unique data pairs after deduplication. We then split the data into training and test sets in a 9:1 ratio and fine-tuned the LLaMA3-8B model using LoRA technology. However, the results were unsatisfactory, with only 42.02% and 15.75% of precursor predictions for solid-state and solution methods, respectively, being identical.

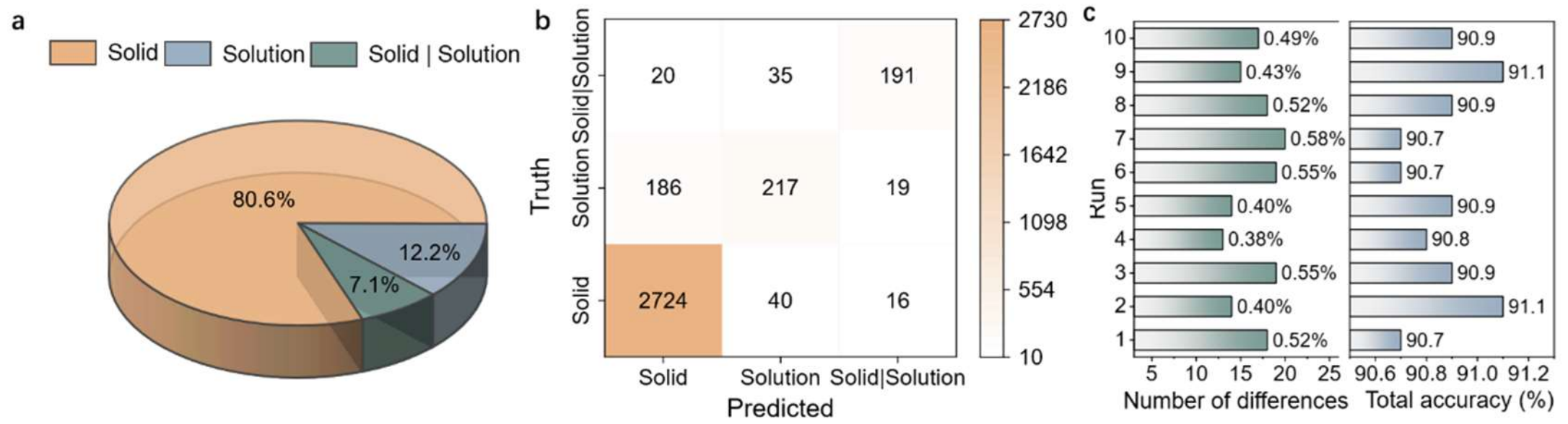

**Figure 3.** **(a)** Proportion of solid-state, solution, and dual-method synthesis among the 31,780

synthesis data points. **(b)** Confusion matrix of the Methods LLM on the test set. **(c)** Variations in results and accuracy of the Methods LLM for 10 additional test set inferences compared to the initial results.

The low accuracy of precursor predictions may be due to insufficient training data. Therefore, we subsequently constructed chemical formula-precursor data pairs, resulting in a total of 31,780 data points after deduplication, which covers most of the elements from 1 to 98 in the periodic table (Figure S4). We then split the data into training and test sets in a 9:1 ratio and re-fine-tuned the LLaMA3-8B model. The accuracy of precursor predictions for the solid-state method reached 80.2%, while for the solution method, it was only 39.7%. Consequently, we decided to focus on predicting precursors for the solid-state method. Thus, we trained another LLM to classify whether a given chemical formula can be synthesized using the solid-state method. This LLM was trained using 31,780 formula-method data points, with chemical formulas synthesized via solid-state, solution, and both methods accounting for 80.5%, 12.2%, and 7.1% of the data, respectively (Figure 3a). The Method LLM achieved an overall classification accuracy of 91.02% on the test set. As shown in Figure 3b, the majority of predictions (93.0%) for solid-state synthesis were accurate. Furthermore, this Methods LLM demonstrated stability over 10 inference runs (Figure 3c), with the largest discrepancy being 18 different results (0.58%) and a maximum accuracy variation of 0.4%.

Next, we conducted a detailed analysis of the performance of the solid-state precursor prediction LLM. In the test set, discrepancies between predicted and actual precursors fell into two cases (Figure 4a): (i) cases of missing or additional precursors (6.7%), and (ii) cases where the number of precursors was correct but included incorrect ones (13.1%). In the first case (Figure 4b), most discrepancies involved either one extra precursor (32.9%) or one missing precursor (63%). In the second case (Figure 4c), the errors included one, two, three, and four incorrect precursors, accounting for 73.5%, 16.4%, 7.1%, and 3.0% of the errors, respectively. Furthermore, we analyzed the above two cases and overall success rates by the number of elements, as shown in Table 1 and Figure S5-10. The success rate of common binary and ternary materials exceeded 91%, with cases of precursor prediction errors below 5%. Moreover, this Precursors LLM passed 10 stability tests, with a maximum discrepancy of only 0.57% in inference results and a maximum success rate variation of 0.3% (Figure 4d). The above analysis demonstrates that the Precursors LLM can provide accurate and robust precursor suggestions, and even when errors or discrepancies occur, they are typically limited to just one precursor in most cases.

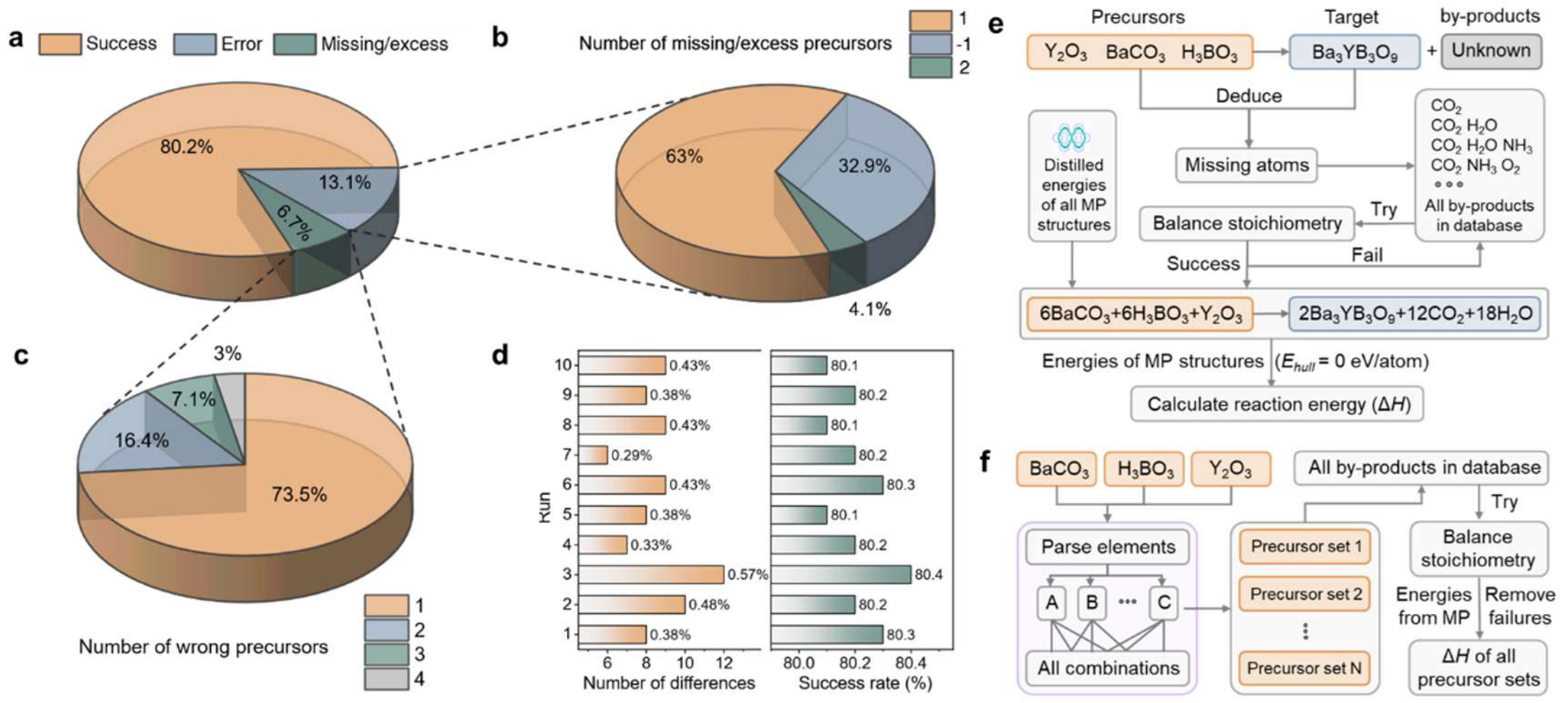

**Figure 4. (a)** Proportions of precursor predictions by the Precursors LLM that are entirely correct, have missing/excess precursors, or contain errors. **(b)** Statistics on the number of cases with missing/excess precursors. **(c)** Statistics on the number of cases with incorrect precursors. **(d)** Variations in results and success rate of the Methods LLM for 10 additional test set inferences compared to the initial results. **(e)** Workflow for inferring byproducts and calculating reaction energies based on precursor predictions from Precursors LLM. **(f)** Workflow for generating alternative precursor sets and calculating reaction energies through elemental permutations and combinations.

**Table 1.** Success rate, error rate and missing/excess rate of the solid-state precursors LLM by number of components.

| Number of components | Success rate | Error rate | Missing/excess rate |
|---|---|---|---|
| all | 80.2% | 13.1% | 6.7% |
| 2 | 92.2% | 4.4% | 3.3% |
| 3 | 91.4% | 4.6% | 4.0% |
| 4 | 76.5% | 14.1% | 9.4% |
| 5 | 68.4% | 24.9% | 6.8% |
| 6 | 56.8% | 37.8% | 5.4% |
| 7 | 50.0% | 33.3% | 16.7% |

By combining the fine-tuned Methods and Precursors LLMs, the precursors of solid-state synthesis can be recommended based on the chemical formulas of materials predicted to be synthesizable by the Synthesizability LLM. To further enhance the reliability and practical applicability of our approach, we calculated the reaction energies before and after synthesis using the MP database and explored alternative precursor sets through permutations and combinations. First, byproducts such as $CO_2$ and $H_2O$ may form alongside the target material during synthesis, which is essential for calculating reaction energies. Therefore, we collected all byproducts from the 31,780 materials synthesis data and identified 30 unique byproduct sets after deduplication. Then the stoichiometry of the reaction equations is balanced by iteratively trying all possible byproducts. Next, the total energies of structures with $E_{hull} = 0$ were extracted from the MP database based on the chemical formulas in the reaction equations to determine the reaction energies. Finally, we identified all possible alternative precursor sets by permuting and combining elements from the initial precursor suggestions (e.g., $Y_2O_3$ could be replaced with $YO_2$ or Y and $O_2$). These combinations were rebalanced and their reaction energies were recalculated. This comprehensive approach broadens the range of potential synthetic routes, making the model more reliable and applicable for real-world synthesis planning.

### 2.4. Identification of synthesizable material structures with outstanding properties

Based on the reliable and accurate Synthesizability, Methods, and Precursors LLMs, we further explored synthesizable functional material structures with exceptional properties. To achieve this, we collected 105,321 theoretical structures from the MP database that lacked experimental synthesis labels and used the Synthesizability LLM to predict their synthesizability. Among these, 45,632 structures were identified as synthesizable. We then compiled data on 27 material properties, including optical, thermodynamic stability, hydration stability, mechanical, optoelectronic, magnetic, topological, thermoelectric, piezoelectric, dielectric, and superconducting properties. For each of these properties, a high-precision graph neural network (GNN) model proposed in previous work, CGTNet, was trained.

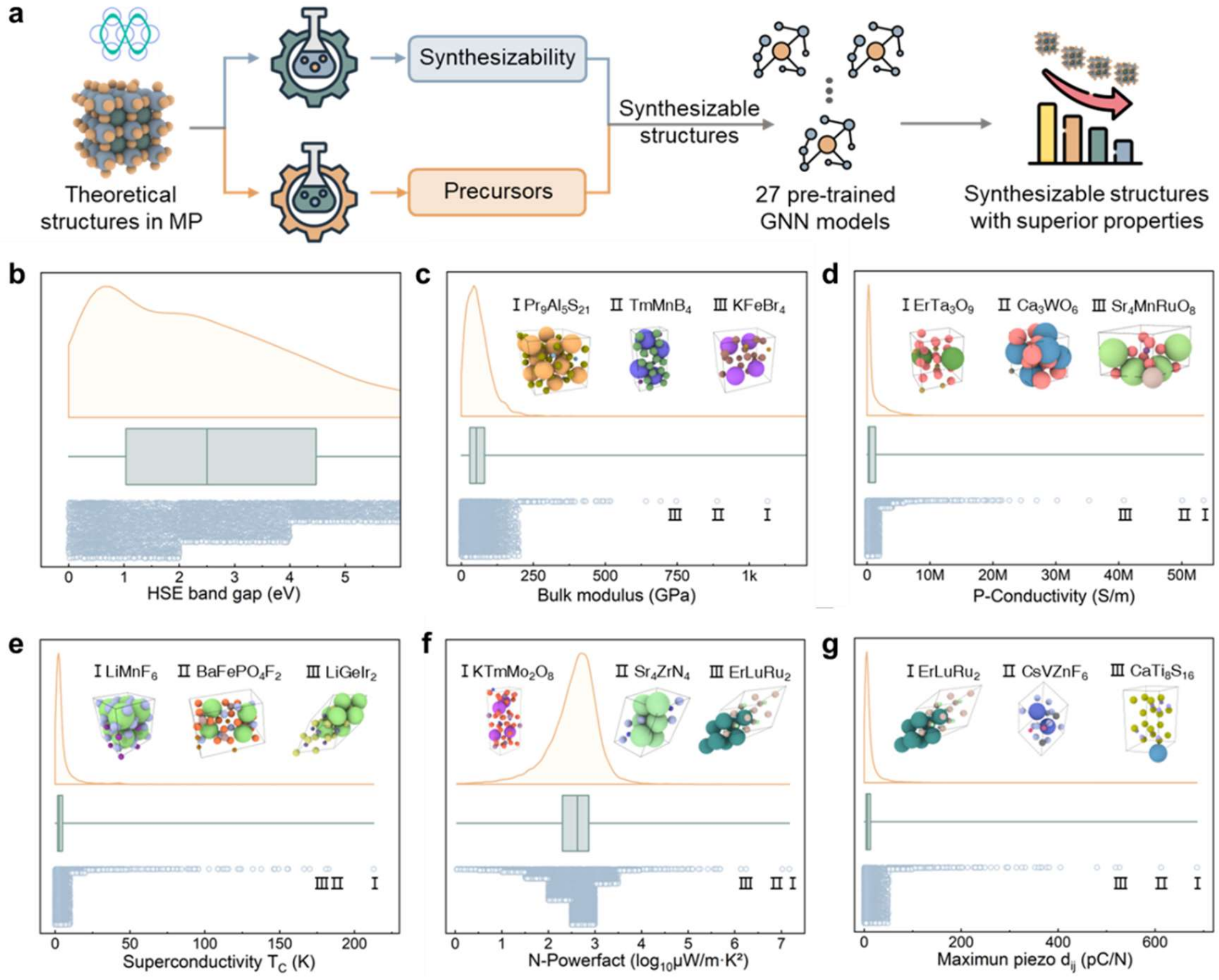

**Figure 5. (a)** Workflow for screening synthesizable structures from theoretical structures in the MP database using the Synthesizability LLM and predicting 27 properties. Distribution, box plot, and jitter plot of predicted properties and the top three material structures for **(b)** HSE band gap, **(c)** bulk modulus, **(d)** p-type conductivity, **(e)** superconductivity transition temperature, **(f)** n-type power factor, and **(g)** maximum piezoelectric coefficient.

**Table 2.** Property name, testing MAE and data amount of 27 GNN model.

| Property | Testing MAE | Data amount |
|---|---|---|
| mBJ band gap | 0.270 eV | 18172 |
| HSE band gap | 0.380 eV | 7376 |
| Experiment band gap | 0.422 eV | 2808 |
| PBE band gap | 0.270 eV | 74992 |
| Electrochemical stability $\Delta G_{pbx}$ | 0.276 eV | 3820 |
| Energy above hull | 0.102 eV | 59635 |
| Bulk modulus | 9.908 GPa | 23824 |
| Shear modulus | 9.933 GPa | 23824 |
| Maximum phonon frequency | 68.14 $cm^{-1}$ | 1265 |
| Exfoliation energy | 37.49 meV/atom | 813 |
| Poisson ratio | 0.087 | 10987 |
| SLME | 4.902 % | 9770 |
| Total magnetic moment | 0.358 A $m^{-2}$ | 74261 |
| Spin-orbit spillage | 0.329 | 11377 |
| Refractive index | 1.27 | 4764 |
| Electronic conductivity | 429208 S/m | 23218 |
| Ionic conductivity | 411463 S/m | 23218 |
| p-Seebeck | 45.295 μV/K | 23218 |
| n-Seebeck | 40.042 μV/K | 23218 |
| p-powerfact | 464.357 μW/cm·$K^2$ | 23218 |
| n-powerfact | 457.080 μW/cm·$K^2$ | 23218 |
| Maximum Piezoelectric $e_{ij}$ | 0.109 C/$m^2$(1og10) | 4799 |
| Maximum Piezoelectric $d_{ij}$ | 0.149 pC/N(1og10) | 3347 |
| Maximum dielectric constant | 0.113 (1og10) | 4809 |
| Maximum electronic dielectric constant | 0.109 (1og10) | 4809 |
| Maximum ionic dielectric constant | 0.199 (1og10) | 4809 |
| Superconducting Tc | 1.764 K | 1057 |

The testing errors (10% data) and data quantities for these 27 property prediction models are shown in Table 2, demonstrating that most models with larger training datasets achieved high accuracy. Consequently, these GNN models were employed to batch predict the properties of the 45,632 synthesizable theoretical structures. The prediction results for six key material properties were visualized: HSE band gap, bulk modulus, p-type conductivity, superconductivity transition temperature, n-type power factor, and maximum piezoelectric coefficient. As shown in Figure 5a, a significant number of synthesizable theoretical structures were identified with HSE band gaps within the semiconductor range (1-3 eV). For the remaining five properties, most synthesizable structures had predicted values that were unremarkable; however, several synthesizable structures with exceptional properties were discovered. The top three structures for each of these properties are highlighted in Figures 5b-g. All predicted properties of synthesizable theoretical structures are shown in Supplementary Data 1.

## Conclusion

In conclusion, we have developed an innovative framework—CSLLM, compromise of Synthesizability, Methods, and Precursors LLM, to predict the synthesizability and precursors of crystal structures. The Synthesizability LLM, fine-tuned with a robust dataset with 140,120 material structures, achieved an ultra-high accuracy of 98.6% on test data, surpassing traditional thermodynamic and kinetic stability methods by 106.1% and 44.5%. Meanwhile, the Methods LLM and Precursors LLM, fine-tuned with 31,780 synthesis data from the literature, demonstrated a high classification accuracy of 91.02% and a success rate of 80.20%. Also, CSLLM incorporates the thermodynamic reaction energy calculations and combinatorial analyses of other potential precursors. Moreover, the user-friendly graphical interface further facilitates practical applications, allowing researchers to upload crystal structures and receive accurate synthesizability and precursor predictions. Owing to the ultra-high prediction accuracy of synthesizability, CSLLM successfully links theoretical material design with experimental synthesis, facilitating the rapid and efficient synthesis of novel functional materials.

## 3. Methods

### 3.1. Fine-tuning and inference of LLMs

Synthesizability LLM and Methods, Precursors LLM are fine-tuned using the pre-trained LLaMA-7B and LLaMA3-8B models, respectively. For Synthesizability LLM, we also tested LLaMA3-8B, achieving a similar overall accuracy of 98.6%. All LLMs were fine-tuned utilizing the LoRA technique [45] with $r = 8$, $\alpha = 32$, and a dropout rate of 0.1. LoRA is designed to efficiently adapt large pre-trained models to downstream tasks by introducing low-rank updates to the model's weights. This approach significantly reduces the number of trainable parameters, making the fine-tuning process more computationally efficient while maintaining high performance. All the fine-tunings and inferences were performed on a server with one NVIDIA A800 GPU with 80GB VRAM and 1TB memory. Notably, the accelerate python library[48] and flash attention[49,50] were used to reduce the GPU memory usage and improve training efficiency.

### 3.2. Training of the CGTNet models

For each of the 27 materials property datasets, we trained our previously proposed transformer-based graph neural network model, CGTNet. This model had demonstrated better accuracy and small data friendliness compared to CGCNN [51], SchNet [52], PaiNN [53], DimeNet++ [54], and GemNet [55]. The main hyperparameters used for training CGTNets are as follows: atom_embedding_size = 256, fc_feat_size = 256, num_fc_layers = 2, gaussians = 50, cutoff = 10.0, out_channels = 512, multi_heads = 2, num_self_att_layers = 2. The meanings of these parameters are explained in Supplementary Note 1.

### 3.3. GUI of CSLLM and Material structure textualization

The GUI of CSLLM was built using the Gradio python library[56]. Uploaded CIF or POSCAR files are automatically converted to material strings. Specifically, the space group of the structure is first analyzed by the SpacegroupAnalyzer class in pymatgen[57] with 'symprec = 0.01' and 'angle_tolerance = 5'. Then, the structure is transformed into its primitive cell, and the Wyckoff positions are analyzed using the PyXtal library[58]. Finally, the material string undergoes inferences through the preloaded Synthesizability, Methods, and Precursors LLM, followed by reaction energy calculations and other precursor recommendations. This entire process runs in the background, and the final output is presented as shown in Figure 1e.

## Acknowledgements

This work was supported by the National Key Research and Development Program of China (grant 2022YFA1503103, 2021YFA1200700), the Natural Science Foundation of China (grant 22033002, 9226111, 22373013, T2321002), and the Basic Research Program of Jiangsu Province (BK20232012, BK20222007). We thank the National Supercomputing Center of Tianjin and the Big Data Computing Center of Southeast University for providing the facility support on the calculations.